\documentclass{PoS}

\usepackage[justification=centering]{caption}
\usepackage{indentfirst}
\usepackage{url}

\title{Study on a prototype of the large dimensional refractive lens for the future large field-of-view IACT}

\ShortTitle{A prototype of the large dimensional refractive lens for the future large FoV IACT}

\author{\speaker{T.L.CHEN}$^{,1,2,3}$,{Z.WANG}$^{2}$,{Q.GAO}$^{1,2}$,{C.LIU}$^{2}$,{Y.ZHANG}$^{2}$,{H.B.HU}$^{2}$,{H.CAI}$^{2}$,{X.Y.ZHANG}$^{2}$,
{H.Y.YANG}$^{4}$,{Y.SHI}$^{3}$,{DANZENGLUOBU}$^{1}$,{M.Y.LIU}$^{1}$,{Z.Y.FENG}$^{2}$,{Y.L.FENG}$^{2}$,{Y.Q.GUO}$^{2}$,
{Q.B.GOU}$^{2}$,{Z.TIAN}$^{2}$ ,{Y.X.XIAO}$^{2}$ \\
        $^1$ Physics Department of Science School, Tibet University, Lhasa 850000, Tibet, China  \\
        $^2$ Key Laboratory of Particle Astrophysics, Institute of High Energy Physics, Chinese Academy of Sciences, Beijing 100049,China  \\
        $^3$ School of Physics and Technology, Wuhan University, Wuhan 430072, Hubei, China\\
        $^4$ Zhao Tong University, Zhaotong 657000,Yunnan, China\\
        E-mail: \email{zwang@ihep.ac.cn}
\thanks{This work was supported by the Research Project of Chinese Ministry of Education (No.213036A), the Grants from the National Natural Science Foundation of China (Nos. 11165013, 11135010, 11105156, 11405180, 11463004, 11375209), the 973 Program of China (No. 2013CB837000) and Youth Innovation Promotion Association, Chinese Academy of Sciences. The speaker T.L.CHEN is also supported by Talents Program for Mount Qomolangma Scholars of Tibet University. We thank Prof. C.Q.SHEN for helpful discussions.}}

\abstract{In gamma ray astronomy, the energy range from sub-100GeV to TeV is crucial due to where there is a gap between space experiments and ground-based ones. In addition, observations in this energy range are expected to provide more details about the high energy emission from GRBs, and thus to understand EBL. Based on the observation results and  the related knowledge, scientists  may be able to unveil the mysteries of galaxy formation and the evolution of early universe. One of the principal issues for next generation Imaging Atmospheric Cherenkov Telescopes (IACT) is to achieve larger field of view (FoV). In this work, we report a refractive water convex lens as light collector to test the feasibility of a new generation of IACT, and some preliminary test results on the optical properties (the focal length, spot size, transmittance, etc.) of a 0.9 m diameter water lens, the photodetectors and DAQ system of a prototype are presented and discussed.}

\FullConference{The 34th International Cosmic Ray Conference,\\
		30 July- 6 August, 2015\\
		The Hague, The Netherlands}

\begin{document}

\section{Introduction}
Larger field of view (FoV) and lower energy threshold are the two most important issues in detecting high energy gamma-ray bursts (GRBs) by ground-based gamma-ray experiments \cite{bib:Inouea,bib:Taboada}. The development of a ground-based technique allowing simultaneous coverage of a significant fraction of the sky is recognised as a high priority issue\cite{bib:Aharonian2001}, two possible approaches have been proposed and attempted in this regard :(i)very large FoV IACT technique based on refractive optics\cite{bib:Lamb,bib:Kifune,bib:Gusumano2006,bib:Gusumano2008} and (ii) dense air shower particle arrays \cite{bib:The ARGO-YBJ Colabration}, or large water Cherenkov detectors installed at very high, 4 km or higher altitude \cite{bib:Hofmann,bib:Sinnis,bib:Westhoff}. On the other hand, reduction of the energy threshold down to sub-100GeV is a critical issue for a number of astrophysical and cosmological problem, e.g. for study of gamma radiation from pulsars and cosmologically distance object like GRBs\cite{bib:Aharonian2004}. It will help not only to improve significantly the flux sensitivities at 100 GeV, but also will open an exciting scientific research area in the intermediate energy between 10-100GeV which connects the space-borne and ground-based gamma-ray astronomy. Presently, very large FoV IACT technique based on refractive optics are still in R$\&$D stage, such as JEM-EUSO and GAW. JEM-EUSO \cite{bib:Kajino} is a Fresnel-optics refractive telescope devoted to the observation of Ultra High Energy Cosmic Rays (UHECR) through the detection of ultra-violet fluorescence light emitted by particles producing showers in the atmosphere. The optics parameters of JEM-EUSO mission are as follows: lens diameter of 2650 mm, FoV of $\pm$ ${30^{\circ}}$, spot size of < 2.8 mm, recently a 1.5 m circular lens system has been tested and its performances have exceeded the requirements \cite{bib:Casolino}. GAW is a new generation of IACT that joins high flux sensitivity with large FoV capability and uses a Fresnel refractive lens as light collector instead of classical reflective mirror \cite{bib:Maccaronea}, and it$^{,}$s optics composed by a custom-made 2.13 m diameter single-sided Fresnel lens, will allow to achieve large FoV capability ( $24^{\circ} \times 24^{\circ}$)\cite{bib:Gusumano2011}.\\
\indent With high transmittance, purified water can be considered as lens medium, therefore we proposed that a refractive water convex lens instead of fresnel lens as light collector to test the feasibility of a new generation of IACT that incorporates high flux sensitivity with large FoV capability, stereoscopic observational approach. Ideal water lens should be in a shape of hemisphere or thick spherical cap, which allows very large FoV and can achieve a uniform image quality on the curved focus surface. In particular, this approach has an obvious  advantage in detecting non-thermal transient phenomena GRBs. More excited is that several GRBs with tens of GeV photos have been detected by space-borne experiments (Fermi/EGRET) \cite{bib:Hurley,bib:Ackermann}. Inspired by above progress, a prototype of large FoV and relative thin water lens has been designed and manufactured in our laboratory. Some preliminary test results on its main characteristics of the prototype are presented and discussed in this paper.\\

\section{Optical properties of water lens}
\subsection{Profile of water lens}

\begin{figure}[!htb]
\centering
\begin{minipage}[t]{0.48\linewidth}
\centerline{\includegraphics[width=5.0cm]{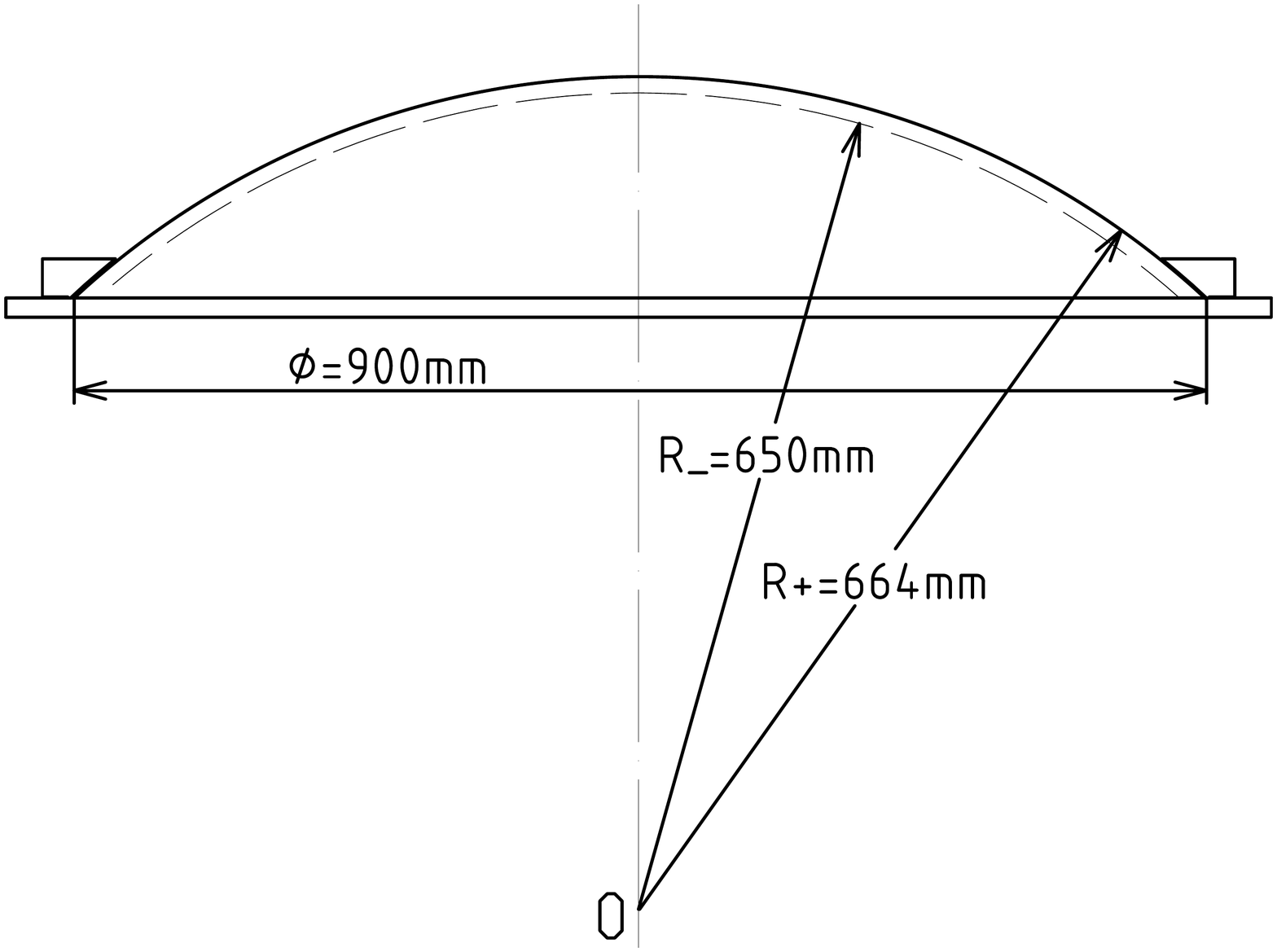}}
\caption{Cutaway view of the water lens.}
\label{cutaway}
\end{minipage}
\begin{minipage}[t]{0.48\linewidth}
\centerline{\includegraphics[width=5.0cm]{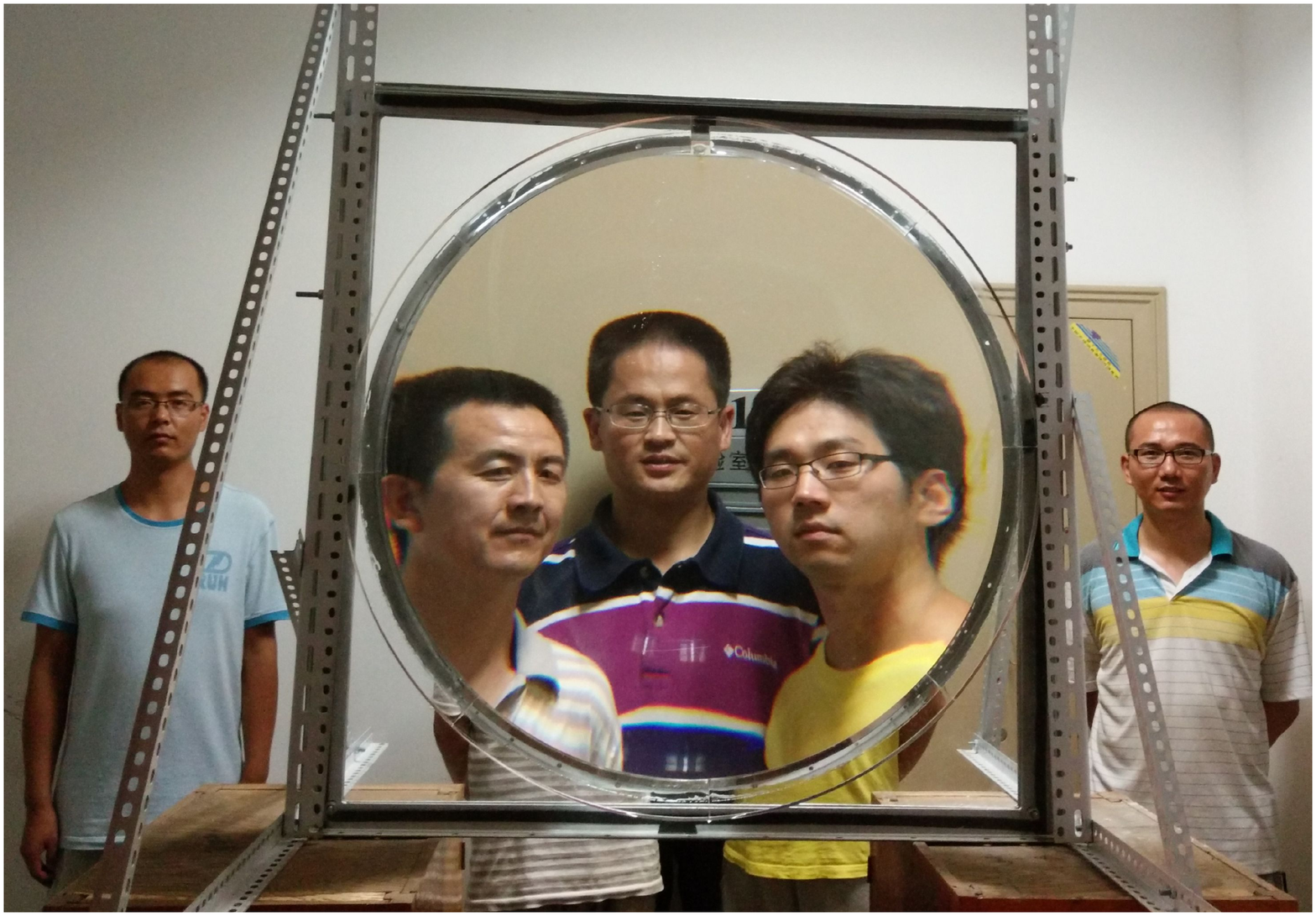}}
\caption{Virtual image formation using \protect\\ the water lens as a magnifying glass.}
\label{lens}
\end{minipage}
\end{figure}

Figure \ref{cutaway} shows the cutaway view of the lens, which is a plano convex,spherical cap lens. The lens is composed of two parts: shell of acrylic and purified water. The main parameters of the lens as follows: diameter is ${\Phi}$=0.9 m, acrylic spherical cap internal and external diameter is  ${R}_{-}$=650 mm and   ${R}_{+}$=664 mm respectively. The average thickness of acrylic layer approximate to ${d}$=14 mm, spherical cap and slab of acrylic layer are glued each other. A tender filling hole was reserved in the edge of the lens to inject purified water. Approximately, the mass of acrylic shell and water is 30 and 70  Kilograms (Kg) respectively.\\
\indent Using the water lens as a magnifying glass, one can see obviously an enlarged, upright, virtual image was formatted from Figure \ref{lens}. \\
\subsection{The focal length and spot size}
\begin{figure}[!htb]
\centering
\includegraphics[width=0.45\textwidth]{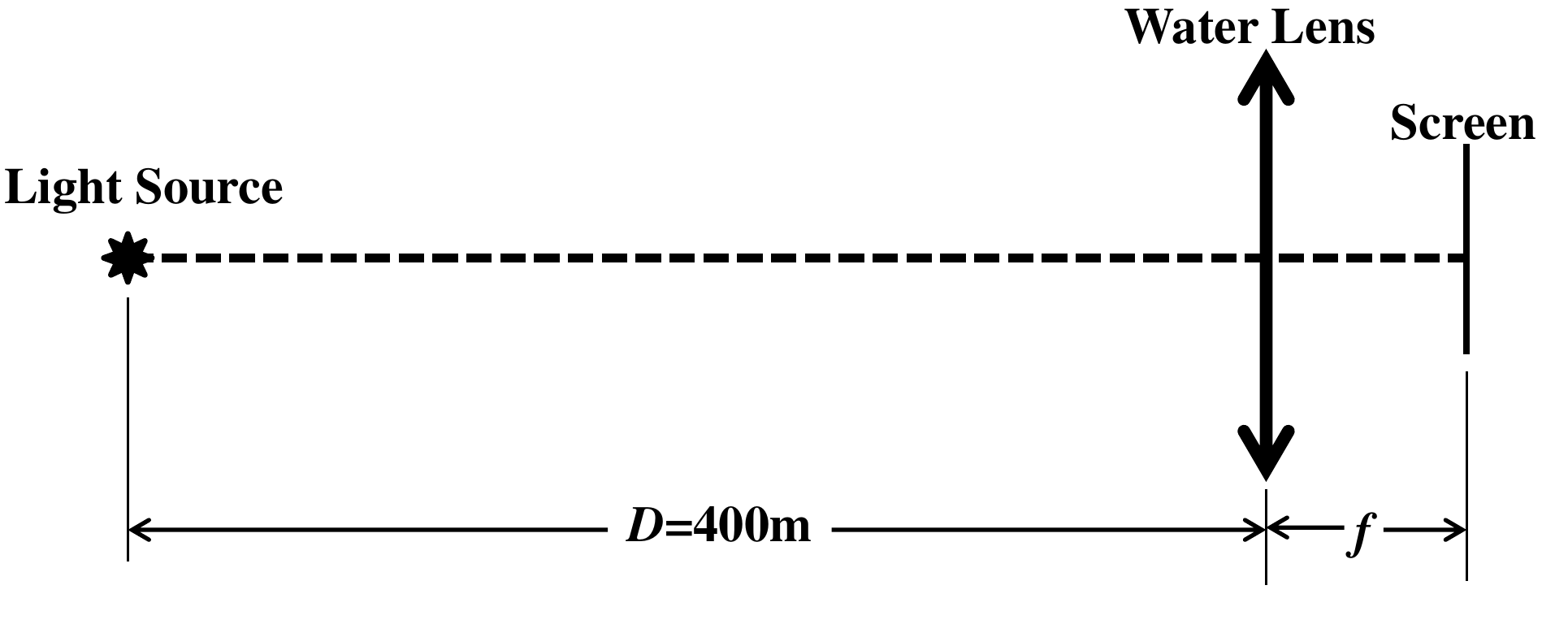}
\caption{The diagrammatic sketch for measuring the FL and spot size.}
\label{focal length}
\end{figure}

The focal length (FL) is one of the most important parameter of a lens. Many methods \cite{bib:Morel}have been elaborated to measure with more or less accuracy a lens$^{,}$ FL.To such large size lens, a traditional focusing method was used to measure the FL. As shown in Figure \ref{focal length} the distance between the light source and the lens was about 400 meters, and we can calculate the angle divergence of light source is only $0.065^{\circ}$, therefore, the beam can be regard as approximate collimated light. A HID (Xenon) Floodlights and headlight of a car were adopted to emit long distance and intense light, the former is a single-chip white LED cup lamp that can provides high lumen output ($\ge$ 350 ${lumen}$). Adjusting the range between the lens and the screen, the optimal image and therefore the FL value can be obtained. As a result, the FL value of measurement is  $f_{exp}$=168$\pm$1.5 cm (Errors are attributed to length measured by tapeline outdoors and the small nonparallel positioning between light source and lens.), and it is basically agreed with simulation result ($f_{sim}$=164 cm) using ZEMAX. The simulation parameters of ZEMAX are as following: lens geometric parameters are taken from the real lens, a collimated light using Photopic (Bright) option in ZEMAX with wave length and respective weight of 470, 510, 555, 610, 650 nm and 0.091, 0.503, 1.000, 0.503, 0.107 (similarly hereinafter). Thus, the ${f}$-number (or focal ratio) ${N}$=${f}$/${d}$=1.8.

\begin{figure}[!htb]
\centering
\begin{minipage}[t]{0.48\linewidth}
\centerline{\includegraphics[width=4.0cm, height=4.0cm]{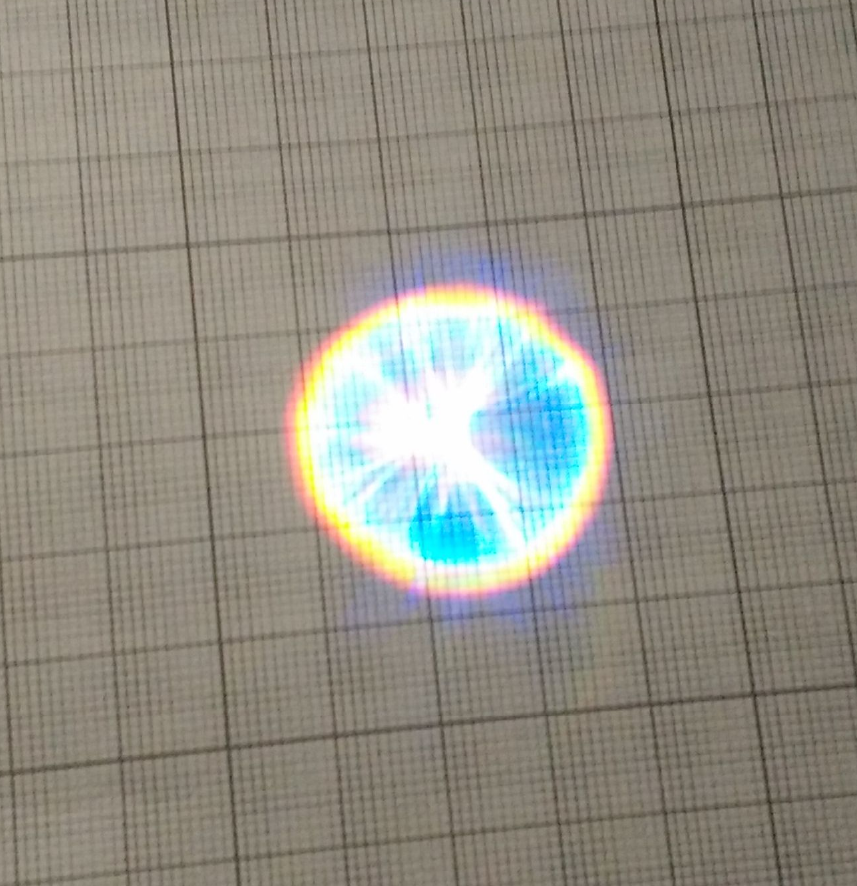}}
\centerline{(a) Photo of image}
\end{minipage}
\begin{minipage}[t]{0.48\linewidth}
\centerline{\includegraphics[width=4.0cm, height=4.0cm]{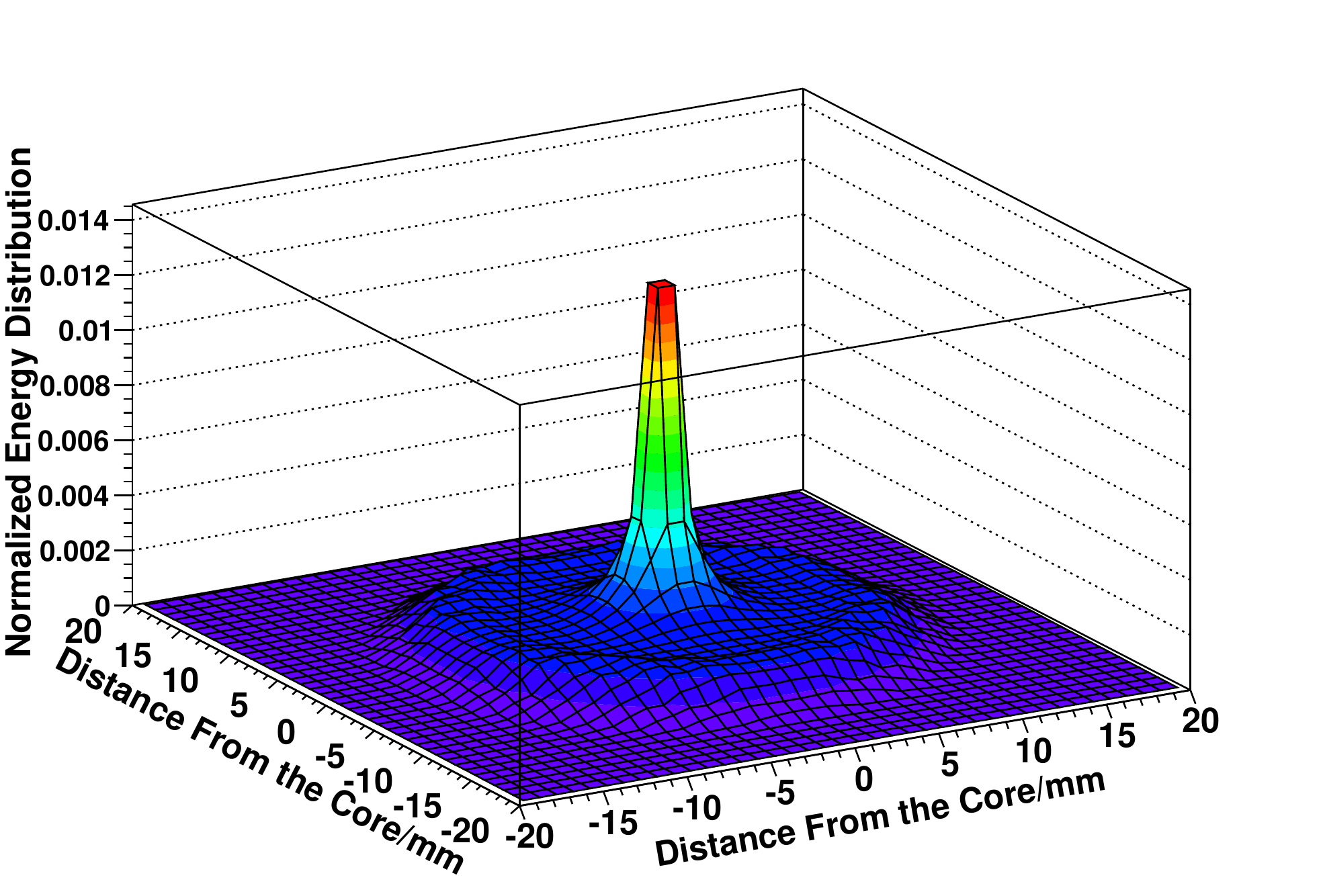}}
\centerline{(b) 2-D simulation result}
\end{minipage}
\caption{(a) Image of HID (Xenon) Floodlights and (b) corresponding simulation results.
}
\label{HID}
\end{figure}

\indent Figure \ref{HID} shows a photo with coordinate paper of the optimal image of HID (Xenon) Floodlights with approximative simulation result which has been calculated using ZEMAX. As can be seen from Figure \ref{HID}(a), the diameter of the focus spot is $D_{expspot}$=32 mm, and it is close to simulation results $D_{simspot}$=30 mm (see Figure \ref{HID}(b)) by using the real FL $f_{exp}$=168 cm in simulation. There is a bit irregular ringlike structures of image, and center is white, middle is blue, outer-ring is yellow,orange and red successively. The structure of image are mainly caused by spherical, chromatic and comatic aberration, which is attributed to the small nonparallel positioning between light source, lens and screen.

\begin{figure}[!htb]
\centering
\begin{minipage}[t]{0.48\linewidth}
\centerline{\includegraphics[width=4.0cm, height=4.0cm]{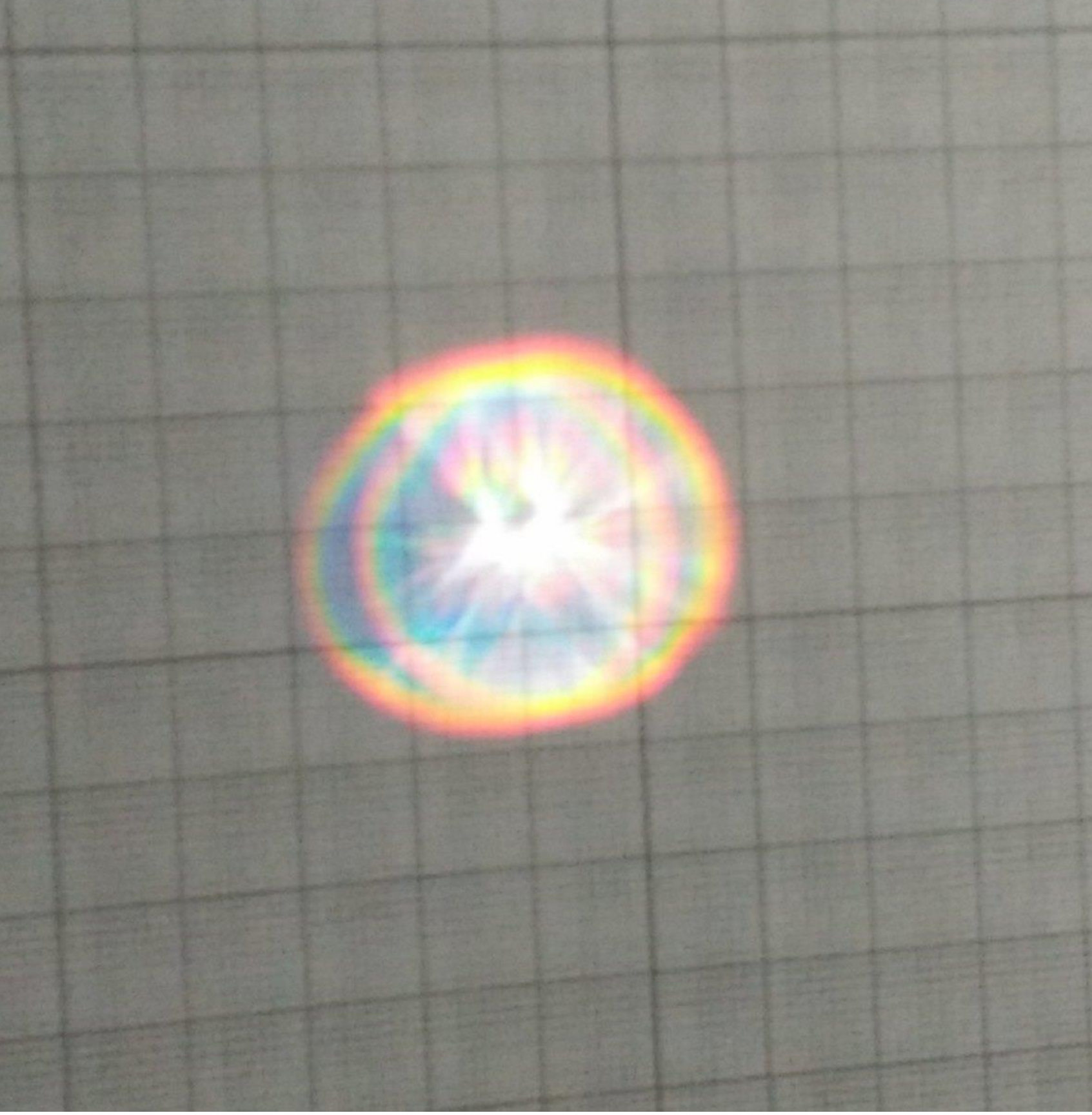}}
\centerline{(a) Photo of image}
\end{minipage}
\begin{minipage}[t]{0.48\linewidth}
\centerline{\includegraphics[width=4.0cm, height=4.0cm]{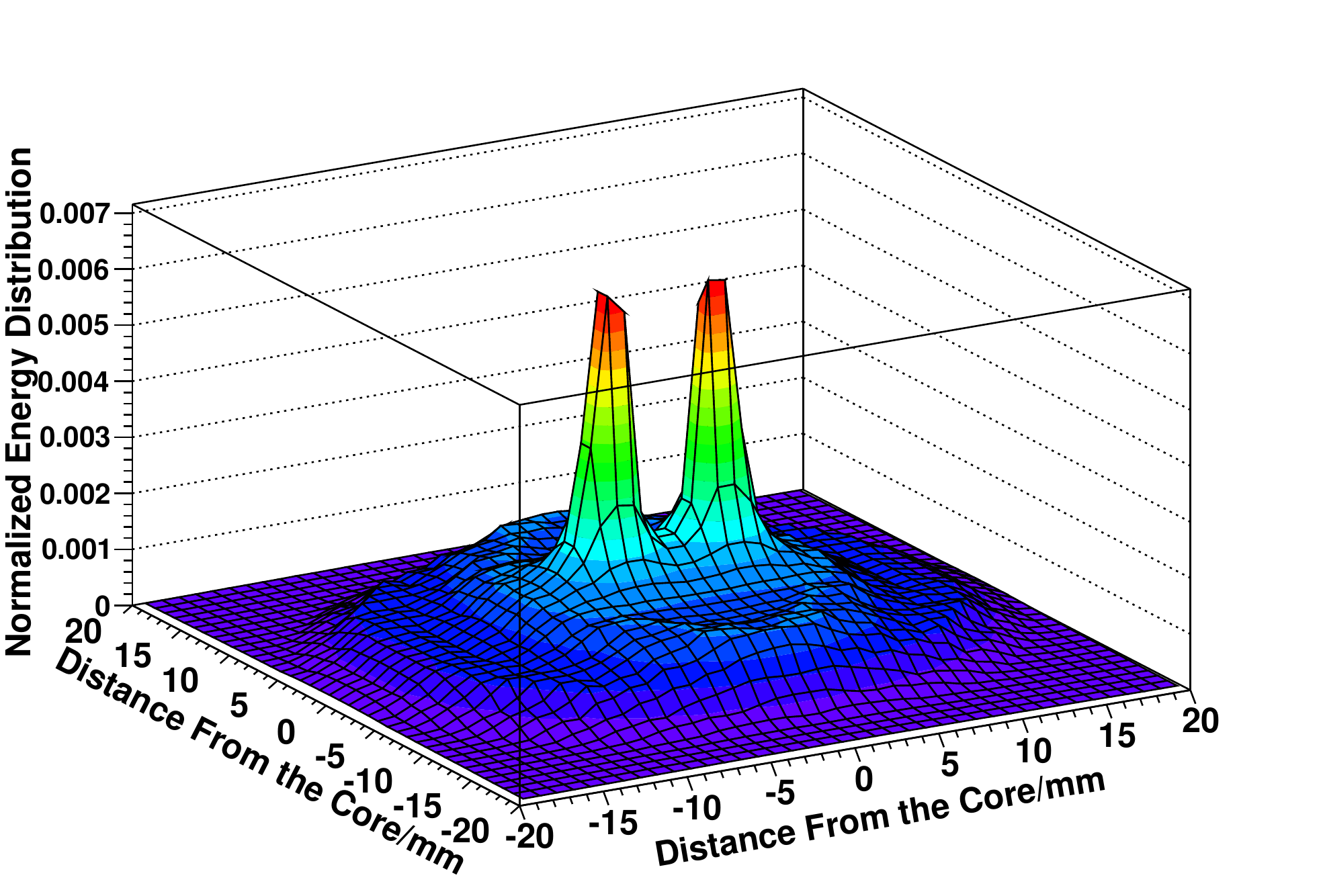}}
\centerline{(b) 2-D simulation}
\end{minipage}
\caption{(a) Image of car headlight and (b) corresponding simulation results.}
\label{car}
\end{figure}

\indent Figure \ref{car}(a) shows a photo of the optimal image of car headlight (distance between two lamps is 1.5 m, we can see obvious superposition of two image. Respective 2-D simulation results are shown in Figure \ref{car}(b) by using the real FL $f_{exp}$=168 cm in simulation.

\subsection{Transmittance}

\begin{figure}[!htb]
\centering
\includegraphics[width=.4\textwidth]{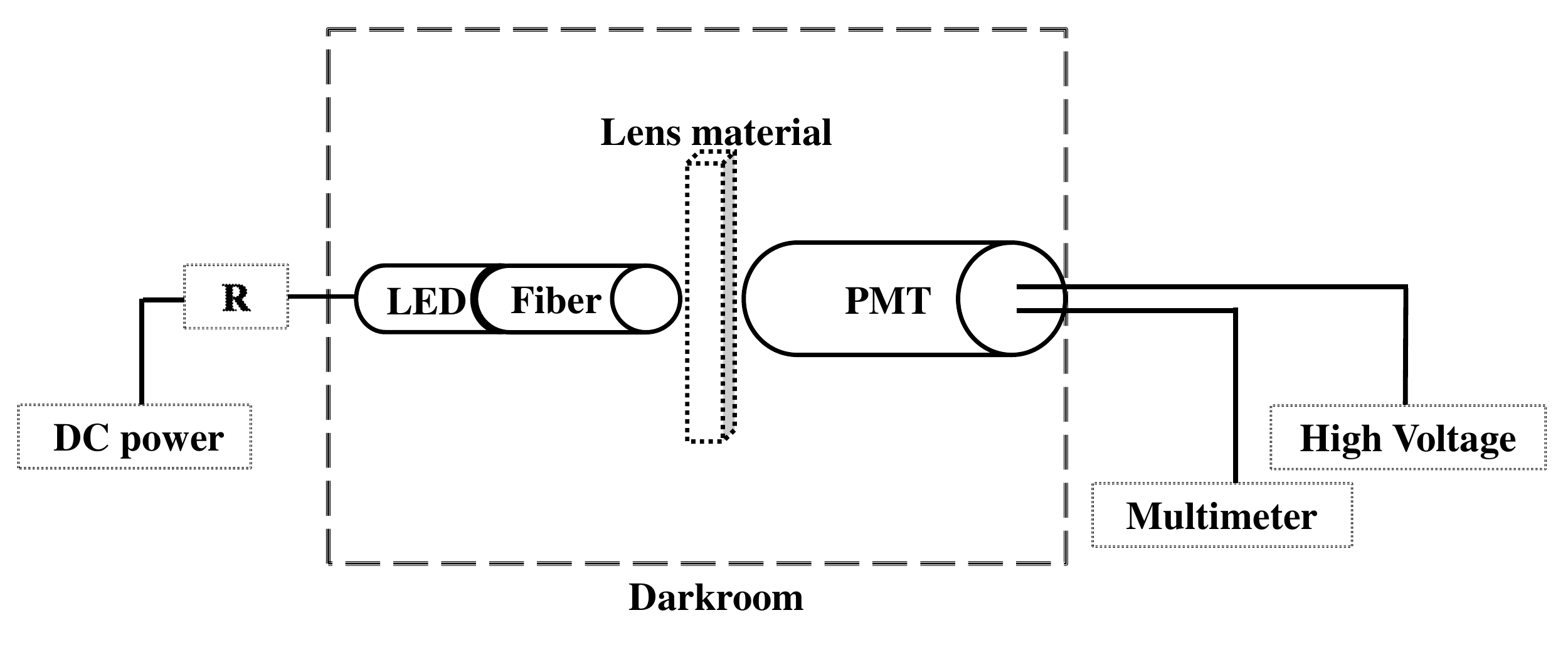}
\caption{Schematic view of direct-current method for measuring material transmittance.}
\label{transmittance}
\end{figure}

As a path-finder experiment, the prototype is designed for exploring the processing technic of large FoV water lens. Hence we chose the commercial material, acrylic, as shell of water lens instead of high UV transparent materials. For detecting the atmospheric Cherenkov light induced by atmospheric shower, we need to know the transmittance of lens which can be inferred from the transmittance of lens shell material generally. A direct-current method was adopted to measure the transmittance of lens shell material, Figure \ref{transmittance} shows the schematic view of the method.

\indent Table \ref{Table I} gives the results of measurements, and we can conclude that the transmittance of single layer material is greater than 93\%, and consequently greater than 86\% of double layer material in the wave length range of 420-660 nm.
\newcommand{\tabincell}[2]{\begin{tabular}{@{}#1@{}}#2\end{tabular}}
\begin{table}[!hbt]
\begin{center}
\caption{Transmittances of lens material versus light wave length}
\label{Table I}
\begin{tabular}{|c|c|c|c|c|c|c|c|c|c|}
\hline
\tabincell{c}{Wave length\\ of LED (nm)} &310&365&420&470&505&525&570&600&660 \\
\hline
\tabincell{c}{Transmittance\\ of measured (\%)} &4.2&5.6&93.3&93.3&94.3&94.6&95.3&93.2&94.7 \\
\hline
\end{tabular}
\end{center}
\end{table}

\subsection{Field of View}

\begin{figure}[!htb]
\centering
\begin{minipage}[t]{0.48\linewidth}
\centerline{\includegraphics[width=5.0cm]{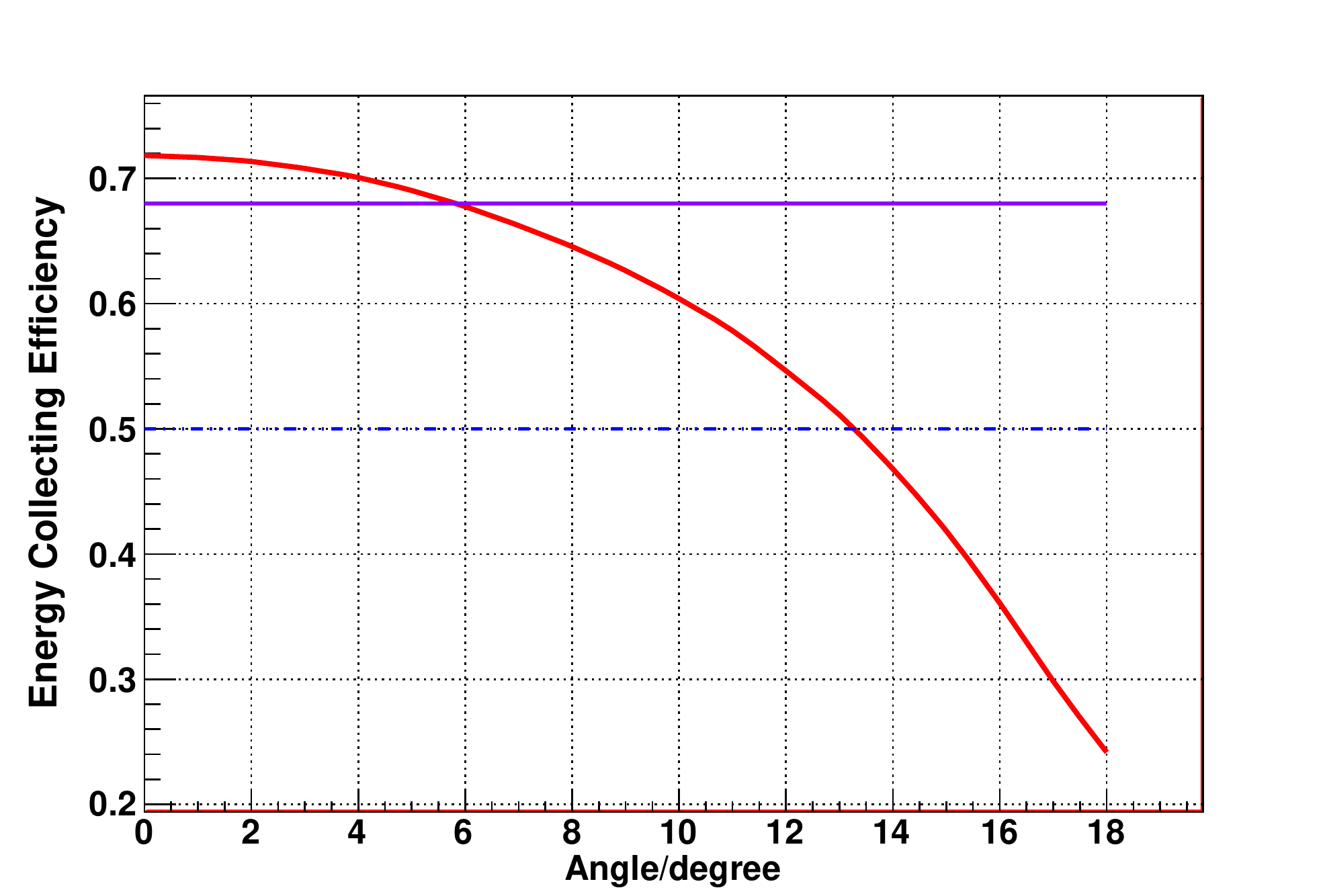}}
\caption{The curve of energy collecting efficiency by a 2 inch PMT on the curved focal plan along with the angle of incidence.}
\label{PMTDetectedRate-new}
\end{minipage}
\begin{minipage}[t]{0.48\linewidth}
\centerline{\includegraphics[width=5.0cm]{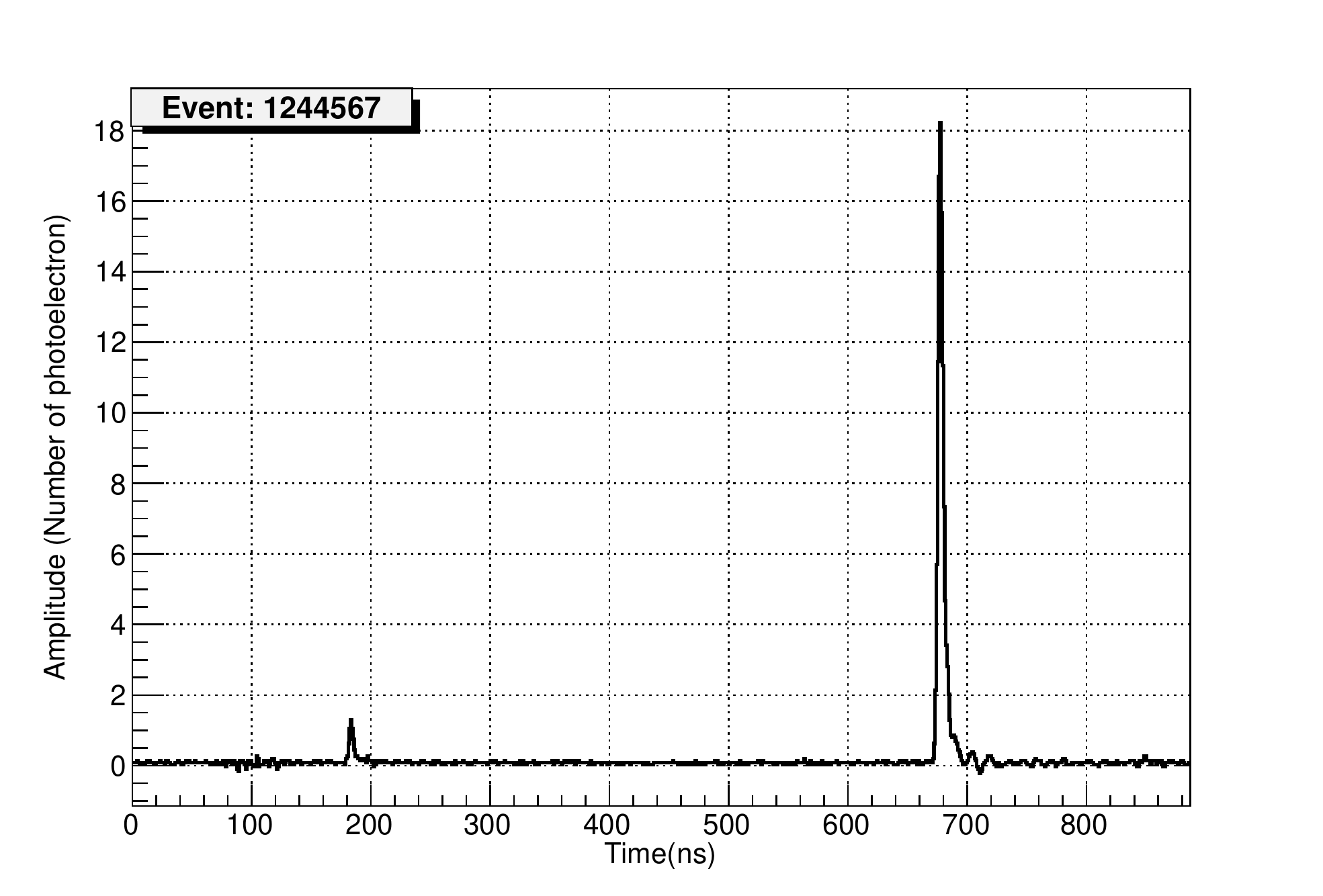}}
\caption{Examples of waveform for events with afterpulse recorded by the FADC.}
\label{afterpulse}
\end{minipage}
\end{figure}

One of the principal parameters of IACT camera is its FoV. For detecting the atmospheric Cherenkov light induced by cosmic rays using the prototype, very preliminary simulation results of the lens show that very large FoV can be achieved. Assuming 80\% of the lens$^{,}$ transmittance and using ZEMAX calculation, the curve of energy collecting efficiency by a 2 inch Photomultiplier Tube (PMT) on the curved focal plan along with the angle of incidence is shown in Figure \ref{PMTDetectedRate-new}. The very preliminary result indicates a FoV of $11^{\circ}\times ^{\circ}$ and $26^{\circ} \times 26^{\circ}$ will be covered in conditions of 68\% and 50\% encircled energy respectively. However, it should be noted that the FoV of the prototype is irrelevant to the thick lens of the final design.

\section{Photodetectors, data acquisition system for prototype}

In order to test the feasibility of the new approach for measuring the atmospheric Cherenkov light induced by gamma ray, firstly, we use the above lens to collect the atmospheric Cherenkov light induced by very high energy (VHE) cosmic rays (CRs). Meanwhile, a 4$\times$4 scintillator Extensive Air Shower (EAS) array will be used to as a coincident measurement with the prototype system. The electronics and data acquisition system (DAQ) of the prototype were built from commercial modules. Mainly, 4$\times$4=16 Hamamatsu R7725 PMT with a 51 mm (2 inch) diameter, bialkali photocathode, 350-600 nm spectral response, head-on type were selected as photodetectors. The PMT signals from the anode were amplified to provide a large dynamic range signals and then they are digitalized by a 12-bit FADC (mod.DT5742\cite{bib:DT5742}, CAEN), which is a 16+1 Channel 12 bit 5 GS/s switched capacitor desktop waveform digitizer. Figure \ref{afterpulse} shows an example of waveform for one event with afterpulse recorded by above-mentioned system, a single photoelectron and a large amplitude afterpulse of 18  NPE (Number of photoelectron) appeared at 190 ns and 780 ns. The afterpulses have  an influence on camera trigger rate\cite{bib:Schlenker}.\\
\indent The telescope will operate at a noise counting rate given mainly by the night sky background. The noise counting rate of the telescope can be calculated knowing the average value of the background and applying the optical and geometrical characteristic quantities of the telescope, with an average value of background rate $N_{bkg}$ of 2200 photons m$^{-2}$ns$^{-1}$s$^{-1}_{r}$ (assuming the Yangbajing International Cosmic Rays Observatory site is the same level with the HESS site \cite{bib:Preu} ), a pixel solid angle $\Delta$$\omega$$_{pix}$ of ${7\times}$$10^{-4}$s$_{r}$, a pupil area ${A}$ of 0.64 m$^{2}$, lens transmission of 80\%, quantum efficiency $\eta$ of 0.2, then the diffuse night sky background (NSB) per PMT is ~1.5 photoelectrons in a time window of 10 ns, that is to say, the rate of NSB noise is 150 MHz. Presumably, we detect the atmospheric Cherenkov light induced by a 100 TeV proton, the rate of event is about ${3.5}\times{10}^{-2}$ Hz and approximate 1000 photoelectrons per nanosecond based on $1\times 10^{-18}$m$^{-2}$s$^{-1}$s$^{-1}_{r}$TeV$^{-1}$  of flux and 16 of PMT, $\Delta$$\omega$${}_{pix}$ of ${7\times}$$10^{-4}$s$_{r}$,100 photos m$^{-2}$TeV$^{-1}$ at Yangbaijng  altitude of Cherenkov light density and the Cherenkov light inside 100 m from the shower core\cite{bib:Aharonian2001}. This and other peculiar characteristics signals constitute the trigger criteria implemented in the trigger logic.\\

\section{Summary}
In order to check the feasibility of a large FoV IACT technique based on refractive optics, a prototype is designed and tested. It is a 0.9 diameter water lens comprised of a plano-convex, thin spherical cap lens. The FL and spot size of the lens are measured which are basically consistent with the simulation results by ZEMAX. Coincidence measurement of the atmospheric Cherenkov light induced by VHE CRs will be achieved with this prototype and a 4¡Á4 scintillator EAS array. Some very preliminary results, such as, the transmittance of shell material and FOV of the lens, the photodetector and DAQ system of the prototype are presented and discussed. \\

\end{document}